\documentclass[%
reprint,
amsmath,amssymb,
aps,longbibliography
]{revtex4-1}
\usepackage{url}
\usepackage[colorlinks=true, linkcolor=blue,urlcolor=blue,anchorcolor=blue,citecolor=blue,bookmarksnumbered]{hyperref}
\usepackage{graphicx}
\usepackage{dcolumn}
\usepackage{bm}

\begin{document}

\title{Functional acoustic metamaterial using shortcut to adiabatic passage in acoustic waveguide couplers}
\author{Shuai Tang$^{1}$}\author{Jin-Lei Wu$^{1}$}\email[]{jinlei$\_$wu@126.com}\author{Cheng L\"u$^{1}$}\author{Jie Song$^{1}$}\author{Yongyuan Jiang$^{1,2,3,4}$}\email[]{jiangyy@hit.edu.cn}

\affiliation{$^{1}$School of Physics, Harbin Institute of Technology, Harbin 150001, China}
\affiliation{$^{2}$Collaborative Innovation Center of Extreme Optics, Shanxi University, Taiyuan 030006, China}
\affiliation{$^{3}$Key Lab of Micro-Optics and Photonic Technology of Heilongjiang Province, Harbin 150001, China}
\affiliation{$^{4}$Key Lab of Micro-Nano Optoelectronic Information System, Ministry of Industry and Information Technology, Harbin 150001, China}

\begin{abstract}
 Shortcut to adiabatic passage~(STAP), initially proposed to accelerate adiabatic quantum state transfer, has been widely explored and applied in quantum optics and integrated optics. Here we bring STAP into the field of acoustics to design compact couplers and functional metamaterial. The space-varying coupling strengths of acoustic waveguides~(WGs) are tailored by means of dressed states in a three-level system, accounting for the desirable acoustic energy transfer among three WGs with short length. We show that the acoustic coupler has one-way feature when loss is introduced into the intermediate WG. More uniquely, when the propagation of acoustic wave is designed to mimic a Hadamard transformation, an acoustic metamaterial can be constructed by arraying several couplers, possessing the beam-splitting function and unidirectional, broadband performances. {Our work bridges STAP and the acoustic coupler as well as metamaterial, which may have profound impacts on exploring quantum technologies for promoting advanced acoustic devices.}
\end{abstract}
\maketitle

\section{Introduction}
Coherent control of quantum states plays a significant role in extensive practical applications including quantum information processing~\cite{Nielsen2010}, high-precision measurement~\cite{Nobel2006}, and manipulation of atoms and molecules~\cite{Kr2007}. To robustly obtain a target state with high fidelity, adiabatic passage~(AP) as well as its variants has been demonstrated as an effective means through well-tuned radiative interactions~\cite{Vit2017,Vitanov_1999,Boozer2008,Unanyan2001,Dridi2009}, while suffering from lengthy evolution owing to the limitation of adiabatic criterion. Recently, methods for speeding up AP, called shortcut to adiabatic passage~(STAP), provide new approaches for rendering a system to reach the desired state quickly, accurately, and robustly~\cite{De2008,Berry2009,XChen2010,Odelin2019}. The development of STAP makes it applicable widely from quantum optics~\cite{BBZhou2017,Vepsalainen2019,TYan2019,JChu2020,YHChen2021} and integrated optics~\cite{Lin:12,Martinez-Garaot:14,HUANG2018187,Chung_2019,Adam2021} to mechanical engineering~\cite{Resines2017,Cercos2020}, physical chemistry~\cite{Vitanov2019,WUJINLEI2019}, and biology~\cite{Weinreich2021,Iram2021}. In this work, we intend to apply STAP for designing compact couplers and functional metamaterial in the field of acoustics.

An acoustic waveguide~(WG) coupler always serves as a functional device to achieve power transfer~\cite{Caowenkang2021,Zhaoshengdong2018,Tangshuai2020,wujinleitangshuai,xuguofu}, mode conversion~\cite{tangshuai2021,Zhuejie2020,QIANJIAO2020}, or phase inversion~\cite{Zhu2016,Zhangjin2020,Zhang2021} by taking advantage of the phononic crystals~\cite{Menchon2014,chenzeguo2016,LIXUEFENG2011}, Helmholtz resonators~\cite{Li_2016,Liuguangsheng2018}, or coiling-up particles~\cite{Ryoo2018,LIYONG2013,Esfahlani2017,Liweibai2020}. Various fascinating phenomena, including focusing beam~\cite{Xie2021,Zhang2009,JIANGYONGYUAN2021,LiuBINGYI2021,TANGAPAC}, splitting beam~\cite{Liweibai2020,Tangshuai2020}, self-bending beam~\cite{Li2021,Tang2021} and vortex beam~\cite{Jiangxue2016prl,Ren2021,Fan2019}, can be observed by constructing acoustic devices integrated by WG couplers. Nevertheless, complicated structural designs pose challenges to modeling and analysis inevitably. Besides, the performance of device is subject to the narrow band and single function, hindering practicability and industrialization of acoustic coupler. {In the past decade, the combination of acoustic systems with other concepts (such as topological state~\cite{Zhangzhiwang2017,Lee2019,Xuehaoran2019,He2020,Xia2018,Wangqiang2021}, parity-time symmetry~\cite{Yangyuzhen2019,Lanjun2020,Fleury2015,ZhuxueFeng2014}, quantum adiabaticity~\cite{Shen2020,Zenglongsheng2021,SHENYAXI2020,Shenyaxi2019} and Landau-Zener transition~\cite{Chenzeguo2021}) has received great interest and has led to surprising advantages for designing advanced functional acoustic metadevices.} As a promising and pragmatic quantum technology, yet STAP in an acoustic system has not been reported so far.

In this work, a technique of STAP is introduced to design compact acoustic WG couplers. Due to the agreement in form between the Schr\"odinger equation in quantum mechanics and the coupled-mode equation of classical waves~\cite{Messiah1962}, the coupled WGs for transporting energy of classical wave can be mapped to quantum states with transitions being driven by external fields. The propagation of acoustic wave along coupled WGs can mimic the quantum evolution controlled with STAP by meticulously modulating discrete coupling strengths between two adjacent WGs. The comparison between performances in couplers designed by STAP and AP verifies the advantage of STAP with a shorter device length. Furthermore, we show that the coupled-WG acoustic system exhibits a one-way transmission feature when introducing an appropriate loss in the intermediate WG. More uniquely, by designing a WG coupler where the propagation of acoustic wave mimics a Hadamard transformation in quantum computing, an asymmetric beam-shaping metamaterial with the beam-splitting function is able to be constructed conveniently by arraying several WG couplers. An incident plane wave can be splitted with a certain angle of refraction in a relatively wide space for one side incidence, but can be hardly transmitted for incidence from the other side. The present work bridges acoustic metamaterials and quantum STAP technology, providing a new solution to design advanced acoustic functional device with excellent performances.
\begin{figure}[b]
	\includegraphics[width=0.88\linewidth]{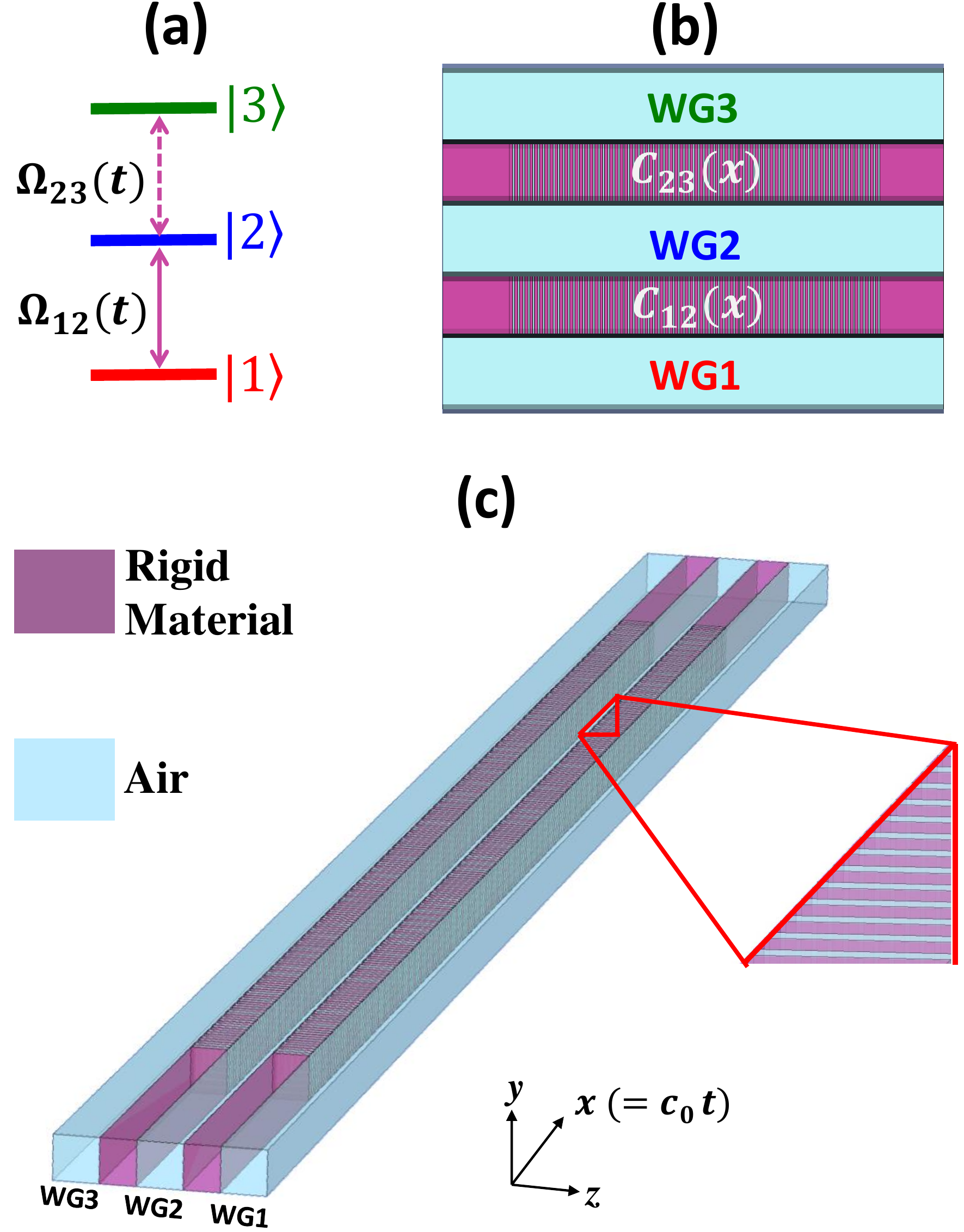}
	\caption{(a)~The three-level quantum system. The states $|1\rangle$, $|2\rangle$, and $|3\rangle$ are coupled by pump field $\Omega_{12}(t)$ and Stokes field $\Omega_{23}(t)$, respectively. (b)~Acoustic three-WG coupler in $y$-$z$ plane. The WGs 1 and 3 are coupled with WG 2 by space-dependent strengths $C_{12}(x)$ and $C_{23}(x)$, respectively. The purple region represents the rigid material and the cyan region represents the air background. (c)~The three dimensional schematic diagram of the acoustic three-WG system.}\label{f1}
\end{figure}
\section{AP in an acoustic system of three coupled WGs}\label{Sec2}
{For a three-level quantum system, as illustrated in Fig.~\ref{f1}(a), ladder-type transitions $|1\rangle\leftrightarrow|2\rangle$ and $|2\rangle\leftrightarrow|3\rangle$ are driven resonantly by two external fields~(for example, visible laser fields for transitions of a valence electron in an atom~\cite{Dudin2021}), respectively, with time-dependent Rabi frequencies $\Omega_{12}(t)$ and $\Omega_{23}(t)$. The Hamiltonian in the interaction picture under the rotating-wave approximation is written as (setting natural unit $\hbar=1$)
\begin{equation}\label{e1}
	\hat{H}(t)=\left(\begin{array}{ccc}
		0 &~\Omega_{1 2}(t) &~0 \\
		\Omega_{1 2}(t) &~0 &~\Omega_{2 3}(t) \\
		0 &~\Omega_{2 3}(t) &~0
	\end{array}\right).
\end{equation}
The two Rabi frequencies can be parametrized by a common amplitude $\Omega(t)$ and a mixed angle $\theta(t)$, as $\Omega_{12}(t)=\Omega(t) \sin \theta(t)$ and $\Omega_{\mathrm{23}}(t)=-\Omega(t) \cos \theta(t)$,
and then instantaneous eigenstates of $\hat{H}(t)$ can be described by a unitary matrix with base vectors $\{|1\rangle,~|2\rangle,~|3\rangle\}$
\begin{equation}\label{e3}
	U_{\mathrm{ad}}=\left(\begin{array}{ccc}
		\sin \theta(t) / \sqrt{2} &~1 / \sqrt{2} &~-\cos \theta(t) / \sqrt{2} \\
		\cos \theta(t) &~0 &~\sin \theta(t) \\
		\sin \theta(t) / \sqrt{2} &~-1 / \sqrt{2} &~-\cos \theta(t) / \sqrt{2}
	\end{array}\right),
\end{equation}
where the first and third rows represent two bright states $\left|\varphi_{\pm}(t)\right\rangle$ with eigenenergies $E_{\pm}(t)=\pm\Omega(t)$, while the second row is called the dark state $\left|\varphi_{0}(t)\right\rangle$ with eigenenergy $E_{0}(t)=0$ (the intermediate state $|2\rangle$ does not get involved in the adiabatic passage process). In the adiabatic basis, the Hamiltonian (\ref{e1}) becomes 
\begin{equation}\label{e4}
	\hat{H}_{\mathrm{ad}}(t)=\Omega(t) \hat{M}_{z}+\dot{\theta}(t) \hat{M}_{y},
\end{equation}
for which three spin-1 operators are introduced
\begin{subequations}
	\begin{equation}
		\hat{M}_{x}=\frac{1}{\sqrt{2}}\left(\begin{array}{rrr}
			0 &-1 &0 \\
			-1 &0 &1 \\
			0 &1 &0
		\end{array}\right),
	\end{equation}
	\begin{equation}
		\hat{M}_{y}=\frac{1}{\sqrt{2}}\left(\begin{array}{rrr}
			0 & i & 0 \\
			-i & 0 & -i \\
			0 & i & 0
		\end{array}\right),
	\end{equation}
	\begin{equation}
		\hat{M}_{z}=\left(\begin{array}{rrr}
			1 & 0 & 0 \\
			0 & 0 & 0 \\
			0 & 0 & -1
		\end{array}\right).
	\end{equation}
\end{subequations}
The term $\dot{\theta}(t) \hat{M}_{y}$ corresponding to nonadiabatic couplings can be ignored in the adiabatic limit $|\dot{\theta}(t)|\ll\sqrt2|\Omega(t)|$~\cite{Kr2007,Bergmann1998,Vit2017}. Then according to the AP technique, a perfect transfer from state $|1\rangle$ to $|3\rangle$ is capable of realizing by utilizing the adiabatic medium state $\left|\varphi_{0}(t)\right\rangle=\cos \theta(t)|1\rangle+\sin \theta(t)|3\rangle$, where the pulse sequence is adopted for ${\theta}(t_i)=0$ and ${\theta}(t_f)=\pi/2$, $t_{i}$ and $t_{f}$ representing the start and end instants, respectively.}

We would like to mimic the quantum system of three discrete levels by constructing an acoustic WG coupler and to achieve desirable energy transfer from an input WG to the target WG. The mathematical similarity of the time-dependent Schr\"odinger equation in quantum mechanics to the wave equation of the classical wave makes it possible to apply the AP technique in the classical-mechanics systems of coupled acoustic WGs~\cite{Shenyaxi2019,SHENYAXI2020,tangshuaiwavelength}. Propagating properties in WGs of acoustic waves make typical ultrafast phenomena in time intuitively visible in space, and also WGs of acoustic waves provide an excellent platform to investigate coherent dynamics accessible difficultly in quantum systems.

We therefore consider an analog of the three-level model in an acoustic system. As shown in Fig.~\ref{f1}(b), three WGs 1, 2 and 3 for an acoustic coupler can be viewed as the discrete states $|1\rangle$, $|2\rangle$, and $|3\rangle$, respectively. The space-dependent coupling strength $C_{mn}$~$(m=1,2;~n=m+1)$ for transmitting sound through slits between WG $m$ and $n$ corresponds to the time-dependent Rabi frequency $\Omega_{mn}$ in the three-level model. Figure~\ref{f1}(c) shows the three dimensional schematic of the acoustic coupler. The width of air pipe (cyan regions) is $w_1$=10 mm. The thickness of rigid material (purple regions) is $w_2$=8 mm. The space-dependent width of slit is $d_{mn}(x)$ between WGs $m$ and $n$. The distance between two spaced slits is $p=4$~mm. The width and height of whole waveguide are $W_0=46$~mm and $h=10$~mm, respectively. By slotting the rigid wall between two adjacent WGs, the WG~1 and WG~3 are coupled with WG~2 when propagating acoustic waves. The coupling strength $C_{12}(x)$ and $C_{23}(x)$ can be modulated along propagation direction ($x$-axis) by adopting different slit widths with relation $C_{mn}(x)\propto d_{mn}(x)$, thus achieving desired space dependence.

Based on the coupled-mode theory~\cite{Messiah1962}, acoustic pressure fields, $P_1$, $P_2$, and $P_3$ satisfying normalization $\sum_{j=1}^{3}\left|P_{j}(x)\right|^{2}=1$, along three WGs in the coupler are described as:
	\begin{equation}\label{e6}
	\begin{aligned}
	&i \frac{\partial P_{1}(x)}{\partial x}=C_{1 2}(x) P_{2}(x), \\
	&i \frac{\partial P_{2}(x)}{\partial x}=C_{1 2}(x) P_{1}(x)+C_{2 3}(x) P_{3}(x), \\
	&i \frac{\partial P_{3}(x)}{\partial x}=C_{2 3}(x) P_{2}(x),
	\end{aligned}
	\end{equation} 
	which can be rewritten as a similar form of Schr\"odinger equation in the spatial dimension
	\begin{equation}\label{e7}
	i \frac{\partial|\varphi(x)\rangle}{\partial x}=\hat{M}(x)|\varphi(x)\rangle,
	\end{equation}
	where $\hat{M}(x)$ is defined as the coupling matrix of the three-WG coupler, corresponding to the Hamiltonian of the three-level system
	\begin{equation}\label{e8}
	\hat{M}(x)=\left(\begin{array}{ccc}
	0 & C_{12}(x) & 0 \\
	C_{12}(x) & 0 & C_{23}(x) \\
	0 & C_{23}(x) & 0
	\end{array}\right).
	\end{equation}
	The acoustic pressure at each site of coupler, $|\varphi(x)\rangle=P_{1}(x)|1\rangle+P_{2}(x)|2\rangle+P_{3}(x)|3\rangle$, can be calculated by solving Eq.~(\ref{e6}). Thereafter, the AP technique for discrete quantum states in a three-level system is capable of mapping into the acoustic three-WG coupler when a spatial adiabatic criterion $|\partial\theta_s(x)/\partial x|\ll\sqrt2|C_s(x)|$ is satisfied, where the parameters for the spatial AP~\cite{Longhi20007,Enrich2016} are introduced by defining $C_{12}(x)=C_s(x)\sin\theta_s(x)$ and $C_{23}(x)=-C_s(x)\cos\theta_s(x)$, and the propagation of acoustic wave could mimic evolution of AP in quantum optics~\cite{Shenyaxi2019}. For the acoustic energy transfer from WG~1~(input port) to WG~3~(output port), one can render the sound distribution in the three WGs to follow the dark state $|\Psi_d(x)=\cos\theta_s(x)|1\rangle+\sin\theta_s(x)|3\rangle$ and set $\theta_s(x_i)=0$ and $\theta_s(x_f)=\pi/2$, where $x_i$ and $x_f$ label the start and end sites of inter-WG slits, respectively.

\section{STAP in the three-WG coupler}\label{Sec3}
\subsection{Modified coupling strengths}
However, a perfect quantum state transfer needs a very long evolution time
to guarantee the temporal adiabatic criterion  $|\dot{\theta}(t)|\ll\sqrt2|\Omega(t)|$, analogous to which an adiabatic acoustic energy transfer with criterion $|\partial\theta_s(x)/\partial x|\ll\sqrt2|C_s(x)|$ requires a long spatial variation of device, costing more space and resources. Therefore, it is necessary to apply a STAP technique to design inter-WG slit widths for achieving acoustic energy transfer in an compact coupler.

{To shorten the length of coupled WGs, we map the time-dependent STAP technique into the space-dependent acoustic system to engineer inter-WG coupling strengths $C_{12}(x)$ and $C_{23}(x)$ as well as the coupling slits by rendering the sound distribution in the three WGs to follow a STAP pathway $|\Psi_s(x)\rangle$,  a dressed state applying for achieving the desired state transfer instead of the AP pathway (i.e., the dark state $|\Psi_d(x)\rangle$):}
	\begin{eqnarray}\label{e08}
	|\Psi_s(t)\rangle&=&\cos \mu_s(x)[\cos \theta_s(x)|1\rangle+\sin \theta_s(x)|3\rangle]\nonumber\\
	&&+i \sin \mu_s(x)|2\rangle,
	\end{eqnarray}
which indicates that two ports of each WG has the same sound distribution as that in the AP situation when the auxiliary parameter $\mu_s(x)$ is limited by $\mu_s(x_i)=\mu_s(x_f)=0$. The STAP technique for a compact WG coupler requires that there are an extra pair of inter-WG coupling strengths $\tilde{C}_{12}(x)$ and $\tilde{C}_{23}(x)$ added into $C_{12}(x)$ and $C_{23}(x)$, respectively, to eliminate the transitions from $|\Psi_s(t)\rangle$ to other orthogonal dressed states, resulting in modified inter-WG space-dependent coupling strengths $C'_{12}(x)=C_{12}(x)+\tilde{C}_{12}(x)$ and $C'_{23}(x)=C_{23}(x)+\tilde{C}_{23}(x)$.
Such a pair of modified coupling strengths can be obtained by several STAP methods, such as invariant-based inverse engineering~\cite{Lewis1969,Chenxi2012}, multiple Schr\"odinger dynamics~\cite{Ibanez2012,XKSong2016}, transitionless evolution engineering~\cite{YHKang2016,YHChen2016}, three-mode parallel paths~\cite{JLWu2017,WUJINLEI2019}, and superadiabatic dressed states~\cite{Baksic2016,BJLiu2017}, which show different mathematical processes but similar underlying physics. In Appendix~\ref{AppendixA}, we give a detailed derivation for obtaining a pair of modified coupling strengths by using the STAP method of dressed states~\cite{Baksic2016} in the context of time-dependent Schr\"odinger equation. Accordingly, a pair of modified inter-WG space-dependent coupling strengths $C'_{12}(x)$ and $C'_{23}(x)$ for the acoustic coupler can be obtained
\begin{eqnarray}\label{e09}
&&C_{12}^{\prime}(x)=-\sin \theta_s(x) \cot \mu_s(x)\partial_x\theta_s(x)- \cos \theta_s(x)\partial_x\mu_s(x), \nonumber\\
&&C_{23}^{\prime}(x)=\cos \theta_s(x) \cot \mu_s(x)\partial_x\theta_s(x)- \sin \theta_s(x)\partial_x\mu_s(x). \nonumber\\
\end{eqnarray}
The related parameters are set as
\begin{equation}
\begin{aligned}
&\theta_s(x)=\frac{\pi x}{2 x_f}-\frac{1}{3} \sin \left(\frac{2 \pi x}{x_f}\right)+\frac{1}{24} \sin \left(\frac{4 \pi x}{x_f}\right),\\
&\mu_s(x)=\frac{\beta_m}{2}\left[1-\cos \left(\frac{2 \pi x}{x_f}\right)\right],
\end{aligned}
\end{equation}
for which we have chosen $x_0=0$.

\subsection{Energy transfer from WG 1 to WG 3}
{To obtain the relation between the inter-WG coupling strength $C$ and the slit width $d$, we consider a two-WG acoustic system composed of WGs~$A$ and $B$ with state function $|\Psi_{T}(x)\rangle=A(x)|A\rangle+B(x)|B\rangle$ and the Hamiltonian $H_{T}(x)=C|A\rangle\langle B|+C|B\rangle\langle A|$, where $A(x)$ and $B(x)$ are the acoustic evolution waveforms along propagation direction in WG~$A$ and WG~$B$, respectively~\cite{shensr,Shenyaxi2019}. Through solving the Schr\"odinger-like coupled-mode equation of the two-WG system $i \frac{\partial}{\partial_{x}}\left|\Psi_{T}(x)\right\rangle=H_{T}\left|\Psi_{T}(x)\right\rangle$, one can obtain $\left|\Psi_{T}(x)\right\rangle=\exp \left(-i H_{T} x\right)\left|\Psi_{T}(0)\right\rangle$ with $\left|\Psi_{T}(0)\right\rangle=A(0)|A\rangle+B(0)|B\rangle$, which yields
\begin{subequations}
	\begin{equation}
		A(x)=\cos (C x) A(0)-i \sin (C x) B(0),
	\end{equation}
\begin{equation}
	B(x)=-i \sin (C x) A(0)+\cos (C x) B(0).
\end{equation}
\end{subequations}}
For a given initial condition of $A(0)=A_0$ and $B(0)=0$, the acoustic power flow along the propagation direction in WG~$A$ and WG~$B$ can be obtained as $P_{\mathrm{A}}(x)={|A(x)|^{2}}/{\left|A_{0}\right|^{2}}=\cos ^{2}(C x)$ and $P_{B}(x)=|B(x)|^{2} /\left|A_{0}\right|^{2}=\sin ^{2}(C x)$.
\begin{figure}
	\includegraphics[width=0.88\linewidth]{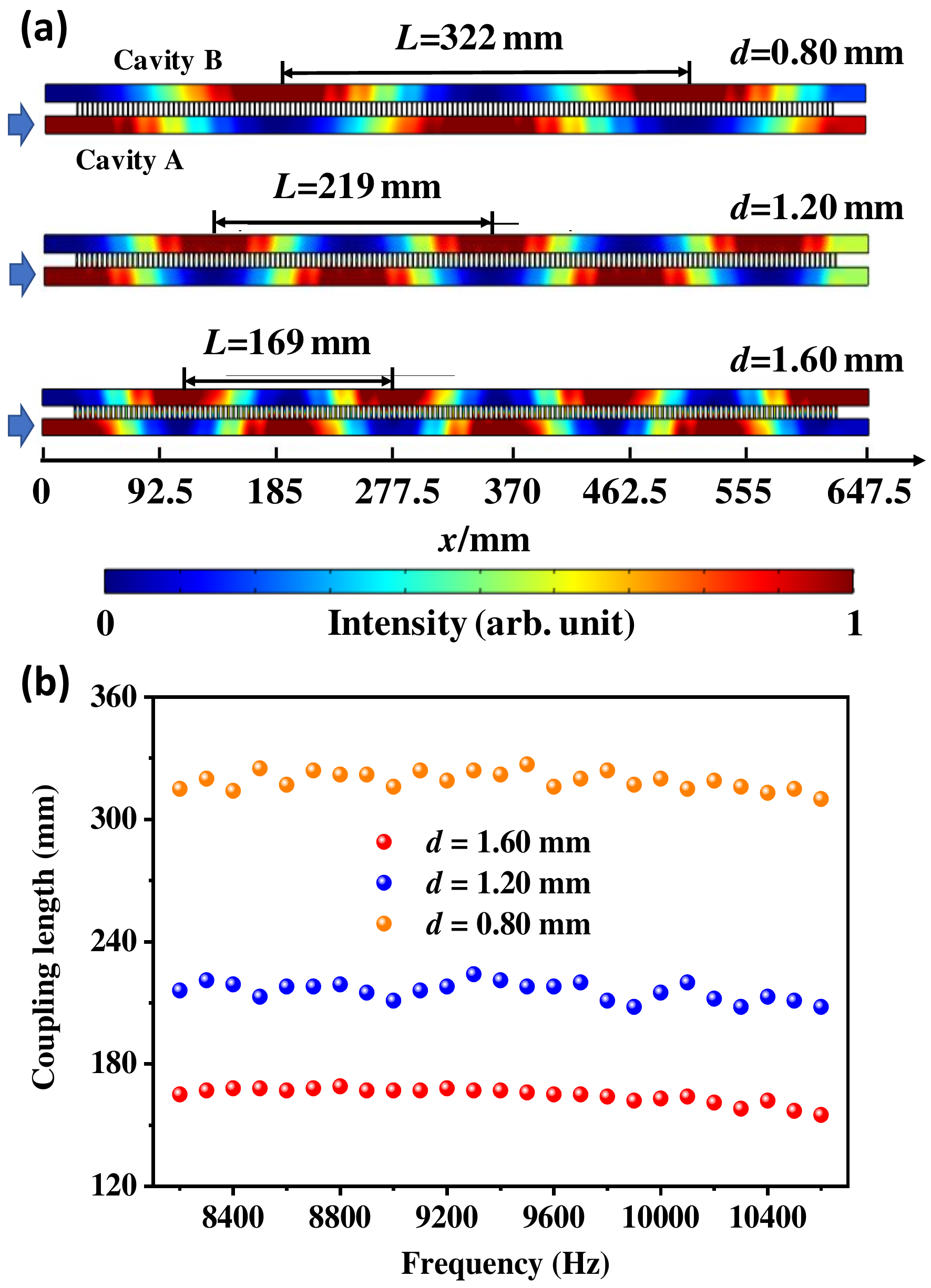}
	\caption{(a)~Acoustic intensity distributions in different acoustic couplers at 8800 Hz, for which the slit widths are $d=0.8$ mm, $d=1.2$ mm and $d=1.6$ mm. Blue arrows represent the input direction and location. (b)~The relation between coupling length and incident frequency.}\label{f2}
\end{figure}
The propagation wave between WGs~$A$ and $B$ will funnel back and forth, leading to a periodic distribution of intensity field. The coupling strength can be calculated by $C=\pi/L$ with $L$ being the coupling length defined by the distance between two adjacent peaks of intensity field. {In a three-WG coupler, the WGs 1 and 3 are coupled with WG 2 by space-dependent strengths $C_{12}^{\prime}(x)$ and $C_{23}^{\prime}(x)$ in Eqs.~(\ref{e09}), respectively. Here, $C_{m n}^{\prime}(x)(m=1,2 ; n=m+1)$ represents the space-dependent coupling strength between WGs $m$ and $n$. To implement the space dependence of coupling strength $C_{m n}^{\prime}(x)$, it is necessary to know how to vary the slit widths between two WGs quantificationally and accurately to change the coupling strengths in space. To this end, it is convenient to obtain the $C$-$d$ correspondence from the results obtained by investigating a two-WG coupler with space-independent slit widths. According to the relation of $C=\pi/L$, we learn that the coupling strength in a two-WG system can be modulated by adopting different values of coupling length. Next, we numerically calculate the coupling length by employing a set of straight two-WG couplers with different slit widths (from 0.2 mm to 2 mm) at 8800 Hz, where the period of slit is set to be $p=4$ mm. Three cases of different slit widths, $d=0.8$ mm, $d=1.2$ mm and $d=1.6$ mm, are shown in Fig.~\ref{f2}(a), while the other cases are listed in Table~\ref{t1}. It can be seen that the coupling length decreases as the slit width increases, and the relation between them can be fitted by the function $L=\alpha/d$ with $\alpha=275 \mathrm{~mm}^{2}$. Therefore, the space-dependent coupling strengths of $C_{m n}^{\prime}(x)$ can be achieved correspondingly by adopting different slit widths along the propagating direction of the coupler. In addition, Fig.~\ref{f2}(b) shows the relation between coupling length and incident frequency for the cases of $d=0.8$ mm, $d=1.2$ mm and $d=1.6$ mm, from which the change of coupling length is very slight within 8200-10600 Hz, providing the foundation for broadband feature of the three-WG coupler.}

\begin{table}
	\caption{{The coupling length ($L$) versus the slit width ($d$)}}
	\centering
	\setlength{\tabcolsep}{5mm}{
		\begin{tabular}{cccc}	
			\hline			
			$d$ (mm)&$L$ (mm)&$d$ (mm)&$L$ (mm)\\
			\hline	
			0.2&1235&0.4&627\\
			0.6&422&0.8&322\\
			1.0&260&1.2&219\\ 
			1.4&191&1.6&169\\
			1.8&151&2.0&142\\
			
			\hline	
			
	\end{tabular}}\label{t1}
	
\end{table}

{The next step is to select an appropriate value of $\beta_m$ to determine the shapes of coupling strengths, so as to determine the required widths of coupling slit along propagation direction by mapping the time-dependent external fields [$\Omega_{12}^{\prime}(t)$ and $\Omega_{23}^{\prime}(t)$] into space-dependent coupling strength [$C_{12}^{\prime}(x)$ and $C_{23}^{\prime}(x)$].
As we illustrated in Appendix~\ref{AppendixA}, the demand for maximum coupling strength decreases when the value of $\beta_m$ increases, offering a variety of options for the design of coupling strengths. Meanwhile, a higher value of $\beta_m$ results in a shorter evolution time (or shorter device length) with a fixed maximum coupling strength [$C_{\max }=\pi /(142 \mathrm{~mm})$ for $d_{\max}=2.0$~mm], indicating that the value of $\beta_m$ is inversely related to the device length. Although a compact acoustic device needs a high value of $\beta_m$ to reduce the length in space, the discrete factor should be taken into account as well because the discrete precision in shorter device is lower than that in a longer one inevitably under a fixed slotting period, which may influence the efficiency of the acoustic energy transfer. Here, we select $\beta_m=0.144\pi$ to ensure sufficient accuracy of discrete slit arrays, from which the device length can be calculated as $L_{1}=316$~mm. Note that we ignore the slits with $d<0.2$~mm at both ends of the device to guarantee the feasibility of fabrication craft in practice, and the actual length of the coupler is $L_{1}=260$~mm in this case.}

\begin{figure}[t]
	\includegraphics[width=1\linewidth]{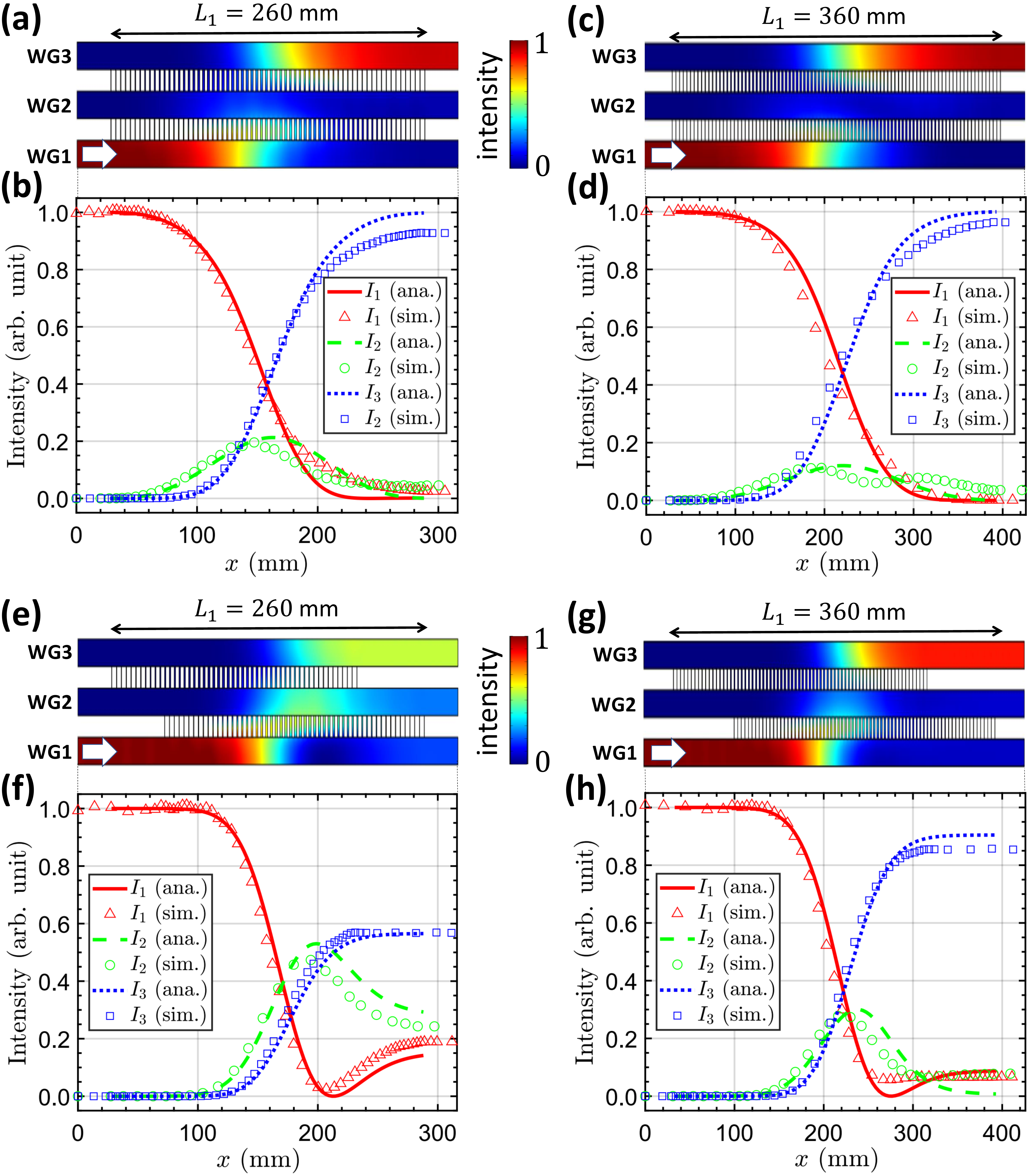}
	\caption{The acoustic fields of the coupler designed by STAP for (a) and (c), {and by~AP with shortcut evolution for (e) and (g).} The acoustic intensity distributions along coupler designed by STAP for (b) and (d), {and by~AP with shortcut evolution for (f) and (h).} White arrows represent the input direction and location. The working frequency is 8800 Hz.}\label{f3}
\end{figure}
To verify the energy transfer feature of the acoustic coupler designed by STAP, the finite element method is adopted to simulate the intensity distribution along each WG at 8800 Hz. The rigid materials are set as hard boundary conditions, and the ports of each WG are set as planar wave radiation conditions. As shown in Fig.~\ref{f3}(a), the acoustic wave input from the left side of WG~1 can transfer to the right side of WG~3 eventually with $L_{1}=260$~mm. The simulated intensity distributions along three WGs shown in Fig.~\ref{f3}(b) agree with the analytical results obtained by solving Schr\"odinger-like equation, Eq.~(\ref{e7}), which confirms the feasibility of our design. It is notable that the small discrepancies between simulated and analytical results are attributed to the discrete precision, which can be diminished by lengthening the device. As shown in Figs.~\ref{f3}(c) and (d), near-perfect energy transfer can be realized as well in the acoustic coupler designed by STAP with actual length of $L_{1}=360$~mm ($\beta_m=0.107\pi$), and the simulated intensity results fit further well with the analytical ones. In addition, the maximum acoustic intensity along WG~2 with $L_{1}=260$~mm is higher than that with $L_{1}=360$~mm, which verifies the conclusion obtained by Eq.~(\ref{e08}) that a higher value of $\beta_m$ leads to a large intensity distribution in WG~2.

\begin{figure}
	\includegraphics[width=\linewidth]{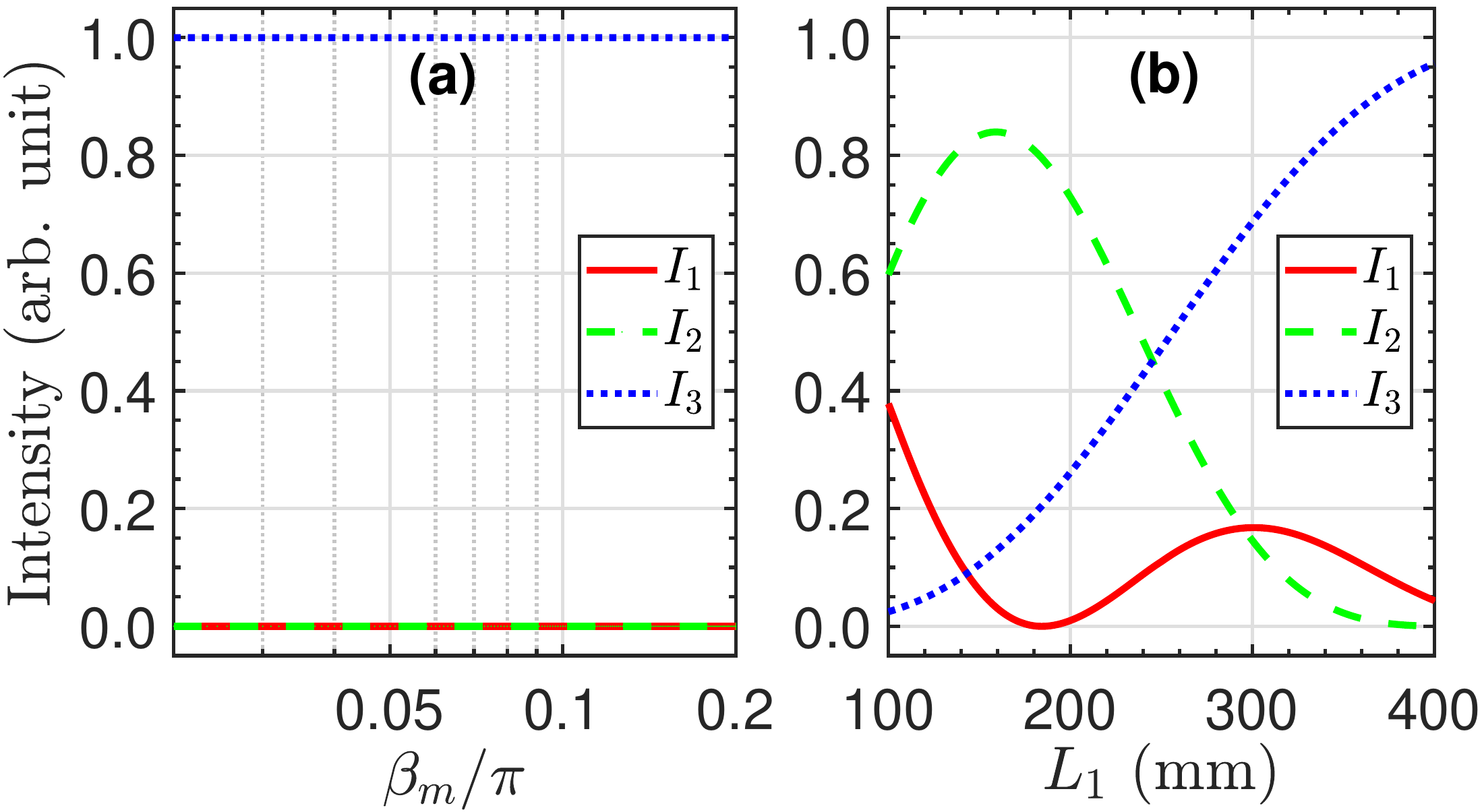}
	\caption{(a)~The relation between the value of $\beta_m$ and the intensity of output wave in acoustic coupler designed by STAP. (b)~Device length-dependent intensity of the output wave in acoustic coupler {designed by AP with shortcut evolution.}}\label{f4}
\end{figure}
\begin{figure}
	\includegraphics[width=1\linewidth]{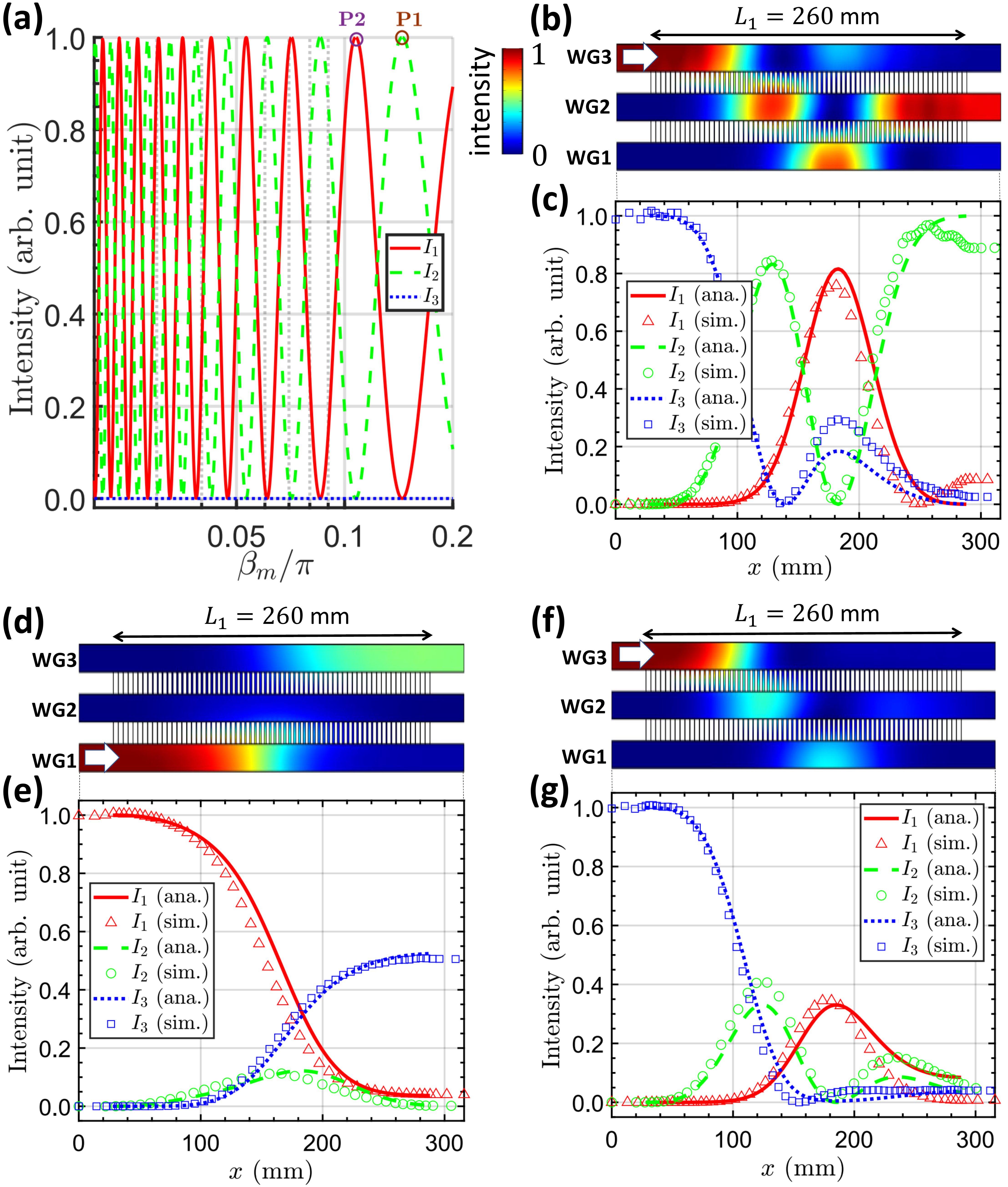}
	\caption{\textcolor{black}{(a)~The relation between the intensity distributions at the output ports of three WGs and the value of $\beta_m$. The acoustic wave is incident from the left side of WG~3 at 8800 Hz. (b)~The acoustic intensity field of the lossless coupler with $L_{1}=260$~mm. (c)~The normalized intensity distributions along three WGs for incidence from WG~3. Acoustic intensity field of the lossy coupler with $L_{1}=260$~mm for incidences from (d)~WG~1 and (f)~WG~3. White arrows represent the input direction and location. The dissipation factor is $\gamma=0.7\max[C'_{12}(x)]$ in WG~2. The normalized intensity distributions along three WGs in lossy coupler for incidence from (e)~WG~1 and (g)~WG~3.}}\label{f5}
\end{figure}

{On the other hand, we show the acoustic coupler designed by AP with shortcut evolution to compare with the performance of the coupler designed by STAP.} As we discussed in Sec.~\ref{Sec2}, a perfect adiabatic transfer from state $|1\rangle$ to $|3\rangle$ is capable of realizing by utilizing the medium state $\left|\varphi_{0}(t)\right\rangle=\cos \theta(t)|1\rangle+\sin \theta(t)|3\rangle$ with ${\theta}(t_i)=0$ and ${\theta}(t_f)=\pi/2$, which can be mapped into space dimension as well~\cite{Shenyaxi2019}. As shown in Ref.~\cite{Shenyaxi2019}, to achieve the asymptotic condition of $\lim _{x \rightarrow 0} \frac{C_{12}(x)}{C_{23}(x)}=1$ and $\lim _{x \rightarrow L_{1}} \frac{C_{12}(x)}{C_{23}(x)}=0$, the space-dependent coupling slits between two adjacent WGs are designed as $d_{12}(x)=d_{0} e^{-(x-x_{i})^{2} / x_{c}^{2}}$ and
$d_{23}(x)=d_{0} e^{-\left(x-x_{i}-\chi\right)^{2} / x_{c}^{2}}$, where $x_{c}=L_{1}/4.8$, $x_{i}=2x_{c}$, $\chi=0.86 x_{c}$, and $d_{0}=343C_{\max}$. When $L_{1}=260$~mm, as illustrated in Fig.~\ref{f3}(e), the acoustic wave inputs from the left side of WG~1 can be transferred partially to the right side of WG~3. Following the simulated and analytical results of normalized intensity distribution shown in Fig.~\ref{f3}(f), {the transfer efficiency in the coupler designed by AP with shortcut evolution is merely around 0.6 due to the imperfect satisfaction of adiabatic condition induced by short device length.} By increasing the length of coupler to $L_{1}=360$~mm, as shown in Figs.~\ref{f3}(g) and (h), the efficiency of  wave transfer is higher than that with $L_{1}=260$~mm but still lower than 0.9, {indicating the drawback of the system designed by AP with shortcut evolution.}

Moreover, in Fig.~\ref{f4}(a) we show the analytic relation between the value of $\beta_m$ and the intensity of output wave from acoustic coupler designed by STAP, where the impact of discrete precision is not taken into account. It can be observed that the normalized intensity of the output wave is independent of the parameter $\beta_m$, and thus the high efficiency wave transfer from WG~1 to WG~3 can be obtained with arbitrary device length so long as the time-dependent driving pulse can be fitted well in the spatial dimension. {For acoustic coupler designed by AP with shortcut evolution}, however, the device length-dependent intensity of the output wave shown in Fig.~\ref{f4}(b) indicates that the transfer efficiency is closely related to the coupler length. Near-perfect energy transfer from WG~1 to WG~2 can be realized only when $L_{1}>400$~mm, which is much longer than that by STAP. Therefore, {compared with traditional AP system with shortcut evolution}~\cite{Shenyaxi2019}, the STAP approach provides an effective way to design compact acoustic device.

\subsection{One-way acoustic energy transfer}\label{Sec3_3}
As another interesting feature, the acoustic coupler designed by STAP can exhibit a one-way energy transfer. If acoustic wave is incident from the left side of WG~3, the corresponding initial condition in STAP system is $|\Psi(t_{i})\rangle=|3\rangle$, and the state function follows the evolution of superposition states of $|\Psi_{\pm}(t)\rangle$ with $a_{+}\left(t_{i}\right)=a_{-}\left(t_{i}\right)=1/\sqrt{2}$. Given by Eq.~(\ref{e13}), the state transfer from $|3\rangle$ to $|2\rangle$ is realized when meeting the same conditions adopted in the state transfer from $|1\rangle$ to $|3\rangle$:
\begin{eqnarray}
		&&\mu\left(t_{i}\right)=\mu\left(t_{f}\right)=0,\quad
		\theta\left(t_{i}\right)=0,\quad
		\theta\left(t_{f}\right)=\pi / 2,\nonumber\\
		&&\beta_{m}=0.144 \pi.
\end{eqnarray}
By mapping the results into spatial dimension, the acoustic wave input from left side of WG~3 can be transferred to the right side of WG~2.

Indeed, the output port for incidence from WG~3 is closely related to $\beta_m$. As numerical results shown in Fig.~\ref{f5}(a), the acoustic output intensity of the propagation wave along WG~3 can be ignored while that along the other two WGs will funnel back and forth with the increase of $\beta_m$. Therefore, either WG~1 or WG~2 can play the role of output port by selecting appropriate value of $\beta_m$. The point marked by P2 on red solid line corresponds to $\beta_m=0.107\pi$ [the device shown in Fig.~\ref{f3}(c)]. The exchange of acoustic energy can be realized in this case because the high-intensity acoustic wave will output on WG~1 (WG~3) with incidence from WG~3 (WG~1). Meanwhile, the point marked by P1 on green dashed line corresponds to $\beta_m=0.144\pi$ [the device shown in Fig.~\ref{f3}(a) and Fig.~\ref{f5}(b)], which illustrates that a high-intensity acoustic wave will output on WG~2 (WG~3) with incidence from WG~3 (WG~1), providing an inspiration for a one-way acoustic energy transfer. The simulated and analytical results of acoustic intensity distribution along three WGs agree well with each other, verifying a port-selective feature of the coupler, i.e., $|1\rangle \rightarrow|3\rangle$~[Fig.~\ref{f3}(a)] and $|3\rangle \rightarrow|2\rangle$~[Fig.~\ref{f5}(b)] .

Taking the acoustic coupler with $L_{1}=260$~mm as an example, the acoustic intensity distribution along WG~2 with incidence from WG~1 [Fig.~\ref{f3}(b)] is quite lower than that from WG~3~[Fig.~\ref{f5}(c)]. Hence, when we introduce a strong wave dissipation in WG~2 with a constant damping rate $\gamma$, the acoustic energy attenuation for incidence from WG~3 will be faster than that from WG~1. In this case, the coupling operator $\hat{M}(x)$ in Eq.~(\ref{e8}) can be changed into a non-Hermitian form
\begin{equation}
\hat{M}_{\gamma}(x)=\left(\begin{array}{ccc}
0 & C_{12}(x) & 0 \\
C_{12}(x)  & -i\gamma & C_{23}(x)  \\
0 & C_{23}(x)  & 0
\end{array}\right).
\end{equation}
The performance of lossy coupler with $L_{1}=260$~mm is illustrated in Figs.~\ref{f5}(d) and (f), from which we can observe that the normalized intensity of output wave for incidence from WG~3 (lower than 0.1) is weaker than that from WG~1 (higher than 0.5). The excellent agreement between simulated and analytical results of the intensity field shown in Figs.~\ref{f5}(e) and (g) further confirms that the output wave for incidence from WG~3 is more sensitive to lossy WG~2 than that from WG~1, and thus the acoustic coupler designed in this case has a one-way wave propagation feature. {Note that the value of $\beta_m$ should be as low as possible to reduce the intensity distribution along WG~2 for incidence from WG~1 so as to guarantee a high contrast ratio between WG~1 and WG~3 incidences, which is opposite to the condition for shortening the device length mentioned before. Therefore, the acoustic coupler designed in our work is a compromise, i.e., the discrete precision, device length, and one-way feature need to be taken into account simultaneously.}

\section{Unidirectional acoustic metamaterial}
\begin{figure}
	\includegraphics[width=1\linewidth]{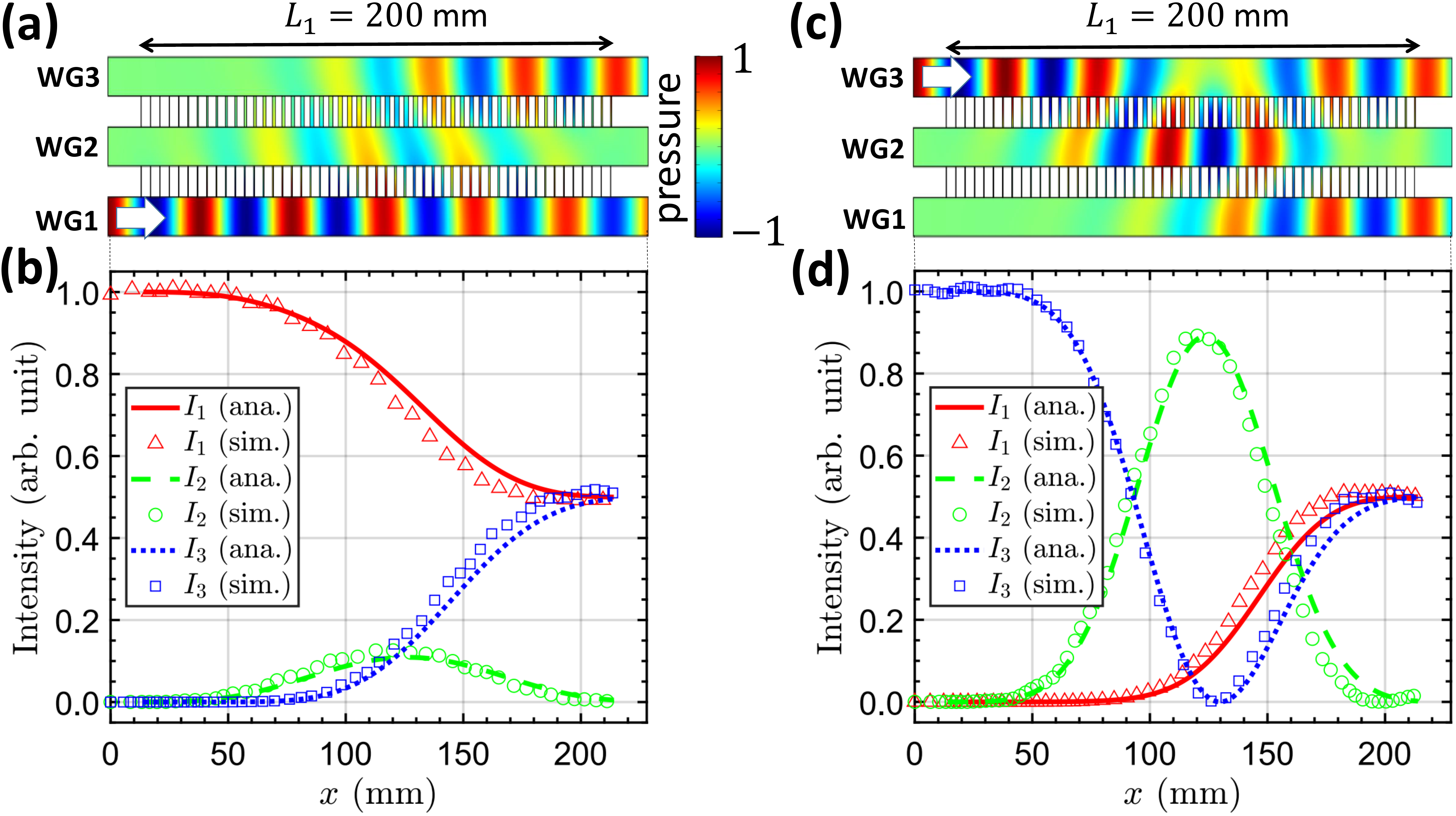}
	\caption{The acoustic pressure field of the coupler with $L_{1}=200$~mm for incidence from (a)~WG~1 and (c)~WG~3. White arrows represent the input direction and location. The normalized intensity distributions along three WGs for incidence from (c)~WG~1 and (d)~WG~3.}\label{f6}
\end{figure}

\begin{figure}[b]
	\includegraphics[width=1\linewidth]{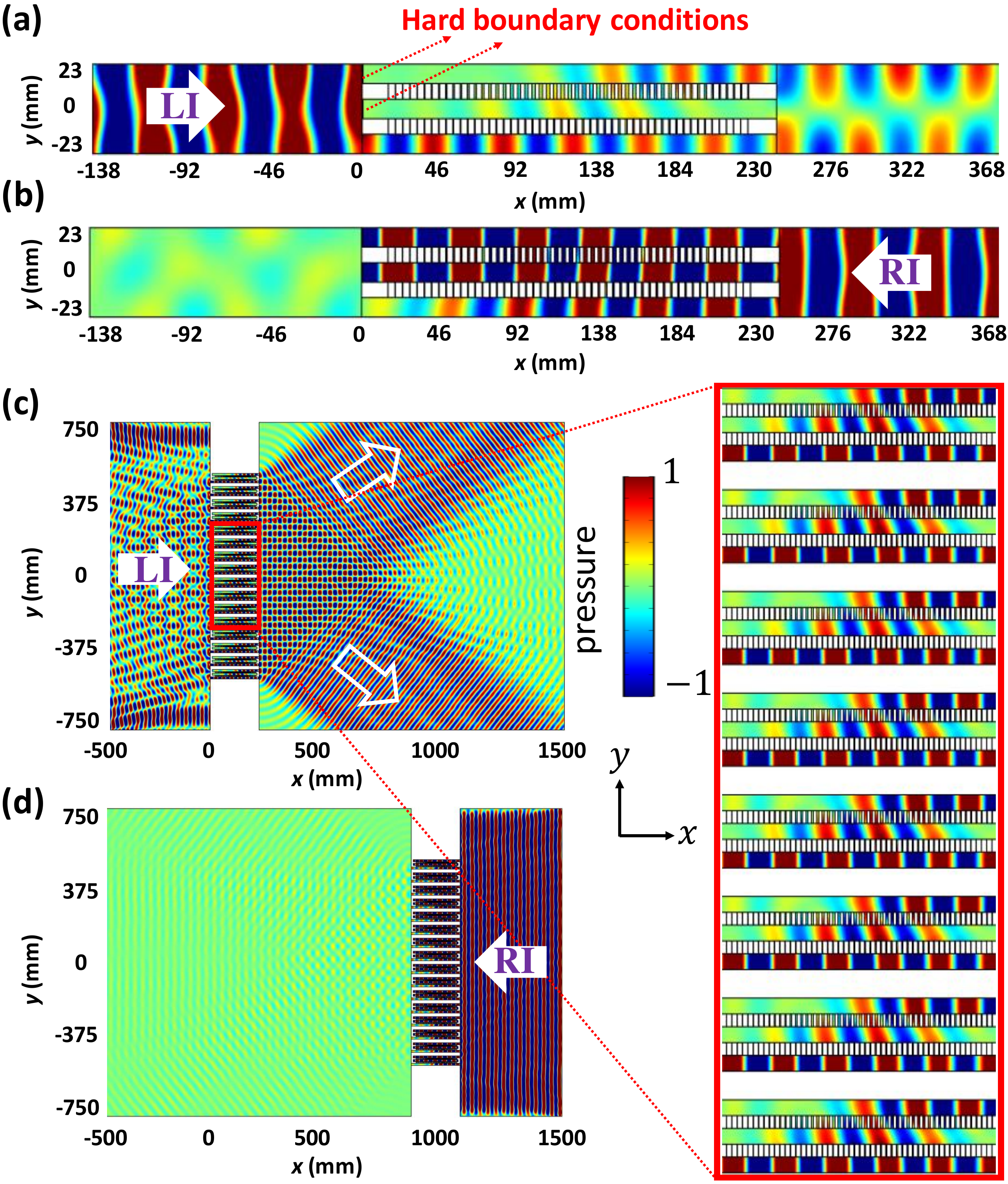}
	\caption{Acoustic pressure fields of the unidirectional mode converter for (a)~left side incidence and (b)~right side incidence at 8800 Hz. Acoustic pressure fields of the unidirectional metamaterial composed of mode converter array for (c)~left side incidence and (d)~right side incidence at 8800 Hz.}\label{f7}
\end{figure}
\begin{figure*}
	\includegraphics[width=0.88\linewidth]{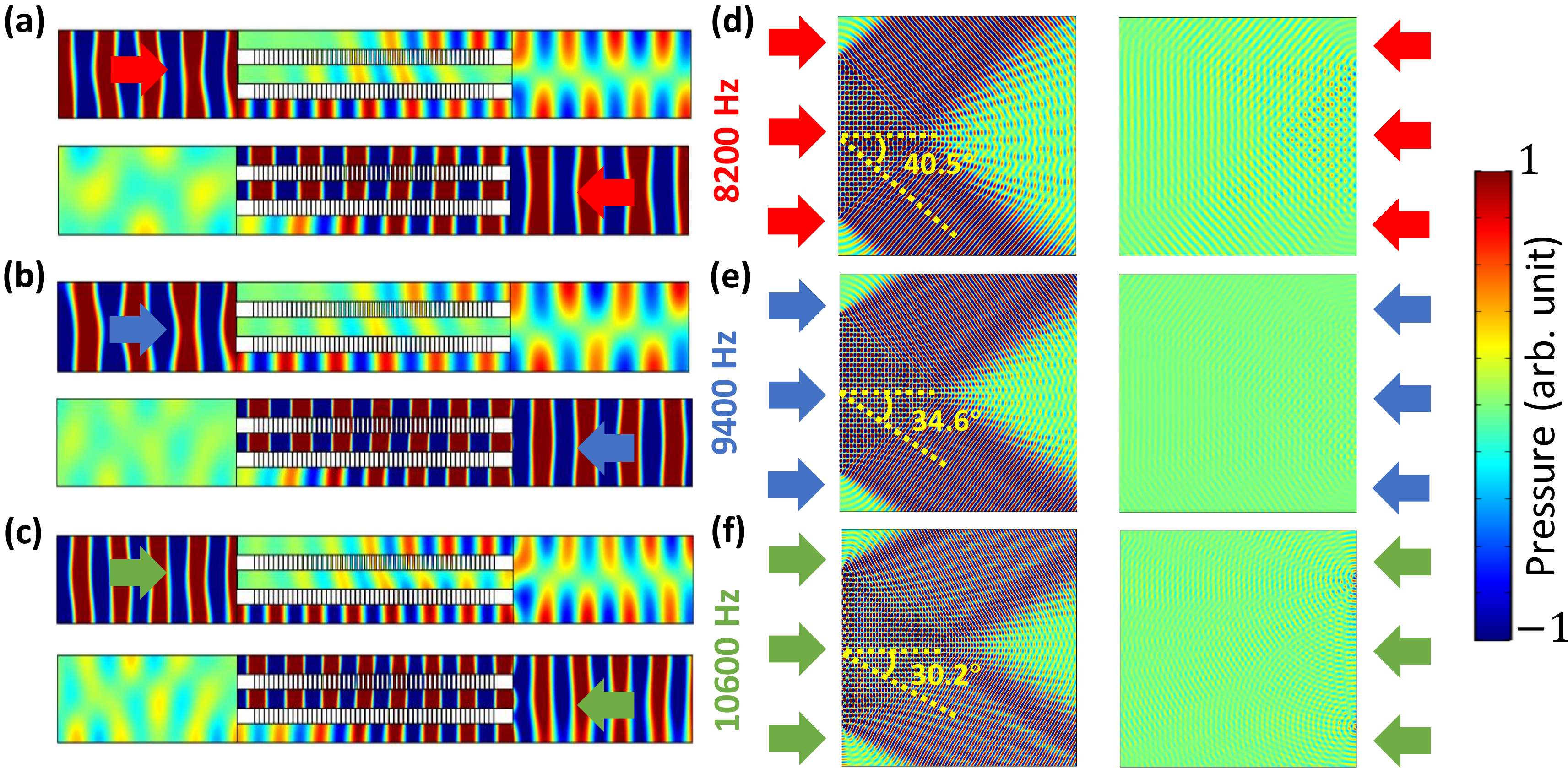}
	\caption{Acoustic pressure fields of the mode converter and metamaterial at (a), (d)~8200 Hz, (b), (e)~9400 Hz, (c), (f)~10600 Hz. The colored arrows represent the incident directions.}\label{f8}
\end{figure*}

In this section, we intend to construct a unidirectional acoustic metamaterial based on the coupler designed by STAP, through which the incident plane acoustic wave can be split in a relatively wide space for one side incidence but hardly transmitted for the other side incidence. A three-WG acoustic coupler that renders a Hadamard-like transformation, $|1\rangle \rightarrow \frac{1}{\sqrt{2}}(|1\rangle-|3\rangle)$ and $|3\rangle \rightarrow\frac{1}{\sqrt{2}}(|1\rangle+|3\rangle)$, is adopted to serve as a basic unit cell to generate desirable output phases for the design of metamaterial. Here, we utilize the conditions
\begin{eqnarray}
&&\mu\left(x_{i}\right)=\mu\left(x_{f}\right)=0,\quad
\theta\left(x_{i}\right)=0,\quad \theta\left(x_{f}\right)=\pi / 4, \nonumber\\
&&\beta_{m}=0.107 \pi
\end{eqnarray} 
to realize the Hadamard-like wave transformation in acoustic coupler, and set $\theta_s(x)=\frac{\pi x}{4 x_{f}}-\frac{1}{3} \sin \left(\frac{\pi x}{x_{f}}\right)+\frac{1}{24} \sin \left(\frac{2 \pi x}{x_{f}}\right)$ and $\mu_s(x)=\frac{\beta_{m}}{2}\left[1-\cos \left(\frac{2 \pi x}{x_{f}}\right)\right]$ for the coupling strengths in Eq.~(\ref{e09}). The performance of acoustic coupler is shown in Figs.~\ref{f6}(a) and (c), where acoustic waves are output from WG~1 and WG~3 simultaneously with identical intensity for a single port incidence from WG~1 or WG~3. The simulated and analytical results of intensity distribution along three WGs shown in Figs.~\ref{f6}(b) and (d) confirm the beam splitting feature of the coupler induced by Hadamard-like wave transformation. Moreover, the phase responses of the output wave from WG~1 and WG~3 are opposite for incidence from WG~1, while those for incidence from WG~3 are the same, providing a fascinating ability to construct unidirectional acoustic device.

Before designing the acoustic metamaterial with beam splitting feature, as shown in Figs.~\ref{f7}(a) and (b), we propose a unidirectional acoustic mode converter constructed by three coupled WGs with the left sides of WG~2 and WG~3 being sealed. For left side incidence (LI), the 0-order plane acoustic wave can only input from the left port of WG~1, and the 1-order wave is able to be generated owing to the opposite phase responses of the output wave from WG~1 and WG~3~\cite{QIANJIAO2020,tangshuai2021}. For right side incidence (RI), however, the reflection field is strong and the transmission field is relatively weak compared with that for LI, which is attributed to that most of propagation waves along WG~1 are coupled into WG~3 and reflected by the hard boundary conditions, showing an excellent performance of unidirectional transmission feature.

Next, we adopt a coding method~\cite{Xie2017} to construct the unidirectional acoustic metamaterial with beam splitting property, where two phase responses, 0 and $\pi$, are needed for structural design, which can be satisfied by output waves from WGs~1 and 3 due to their phase difference of $\pi$. By arraying the mode converter periodically along $y$-direction, as shown in Fig.~\ref{f7} (c), a beam splitting metamaterial is obtained, and the pattern of transmission field is determined by the period length ($L_{p}$) and the incident wavelength ($\lambda$). For LI, the diffraction of $\pm$1-order wave are able to directly take one-pass propagation following the round-trip process demonstrated in the work~\cite{Fu2019}, which is preferential to the 0-order wave, thus resulting in a splitting beam. Given the generalized Snell's law~\cite{Yu2011}, the refraction angle can be calculated by $\theta_{r}=\arcsin (\lambda / L_{p})$, which makes it necessary to guarantee the relation of $L_{p}>\lambda$. Otherwise, surface acoustic wave will be produced along $y$-direction~\cite{meijunnjp2014} and the desirable high-intensity splitting beam is not able to be obtained in this case. To avoid the generation of evanescent wave, we set $L_{p}=0.064$ m, which is longer than wavelength $\lambda=(343~\mathrm{m/s})/(8800~\mathrm{Hz})=0.039~\mathrm{m}$. As illustrated in Fig.~\ref{f7} (c), the normally incident plane wave can radiate into two directions with $\theta_{r}=\pm37.5^{\circ}$, showing an excellent performance of beam splitting. For RI, however, the intensity of transmission field is weak owing to the strong reflection induced by mode converter array. As illustrated in Fig.~\ref{f7} (d), no apparent beam splitting can be observed, which confirms the unidirectional feature of the acoustic metamaterial proposed in our work. 

Note that there are numerous achievements in the field of acoustic metamaterial so far. From the perspective of structural design, however, traditional approaches mainly utilize complicated unit design with resonance feature to achieve desirable phase responses, which result in unavoidable complexity of modeling and analysis. By taking advantage of the Hadamard-like acoustic transformation, our work provides a new solution to construct acoustic metamaterial with a simple means (slotting the rigid material periodically). Furthermore, the dependence on resonance response makes the bandwidth being a limitation for application of the traditional device, while the acoustic device designed by STAP relies on the coupling effect rather than resonance effect. Consequently, an excellent robustness to the incident wavelength can be realized here. {The broadband feature of the three-WG coupler has been illustrated in Appendix~\ref{AppendixB}.} Therefore, as shown in Figs.~\ref{f8}(a)-(f), the unidirectional acoustic mode converter and metamaterial can work well at 8200 Hz, 9400 Hz and 10600 Hz, verifying the broadband feature of our design. {Meanwhile, the refraction angle marked in Figs.~\ref{f8}(d)-(f) for LI reduces with the increase of working frequency following the relation of $\theta_{r}=\arcsin (\lambda / L_{p})$, which indicates that the incident wavelength can be regarded as a degree of freedom to modulate the angle of  splitting beam dynamically.}

\section{Conclusion}
An approach is exhibited in this work for efficient transfer of acoustic wave energy by connecting shortcut to adiabatic passage (STAP) with the construction of a compact acoustic coupler, expanding applications of STAP into the field of acoustics. The acoustic wave propagating along the coupler mimics the evolution of STAP for a quantum three-level system by modulating the coupling strength between two adjacent waveguides (WGs), thus resulting in a desirable behavior of energy exchange with one-way feature. Meanwhile, we further illustrate that a unidirectional acoustic mode converter is able to be designed via a three-WG coupler rendering a Hadamard-like wave transformation. Thereafter, a broadband acoustic metamaterial with simultaneous unidirectional transmission feature and beam splitting property can be designed by following the generalized Snell's law through periodically arraying the acoustic mode converter, which improves the practicability of the device in engineering applications including sound isolation, communication and stealth. Our work, by introducing STAP into constructing acoustic coupler as well as acoustic metamaterial, paves a new way to guide the design of advanced acoustic functional device from the behind physical laws.
\section*{Acknowledgements}
The authors acknowledge the financial support by the National Natural Science Foundation of China (Grant No. 62075048) and Natural Science Foundation of Shandong Province of China (Grant No. ZR2020MF129).

\appendix

\section{Time-dependent STAP method of dressed states in a quantum three-level system}\label{AppendixA}
\begin{figure}[b]
	\includegraphics[width=0.88\linewidth]{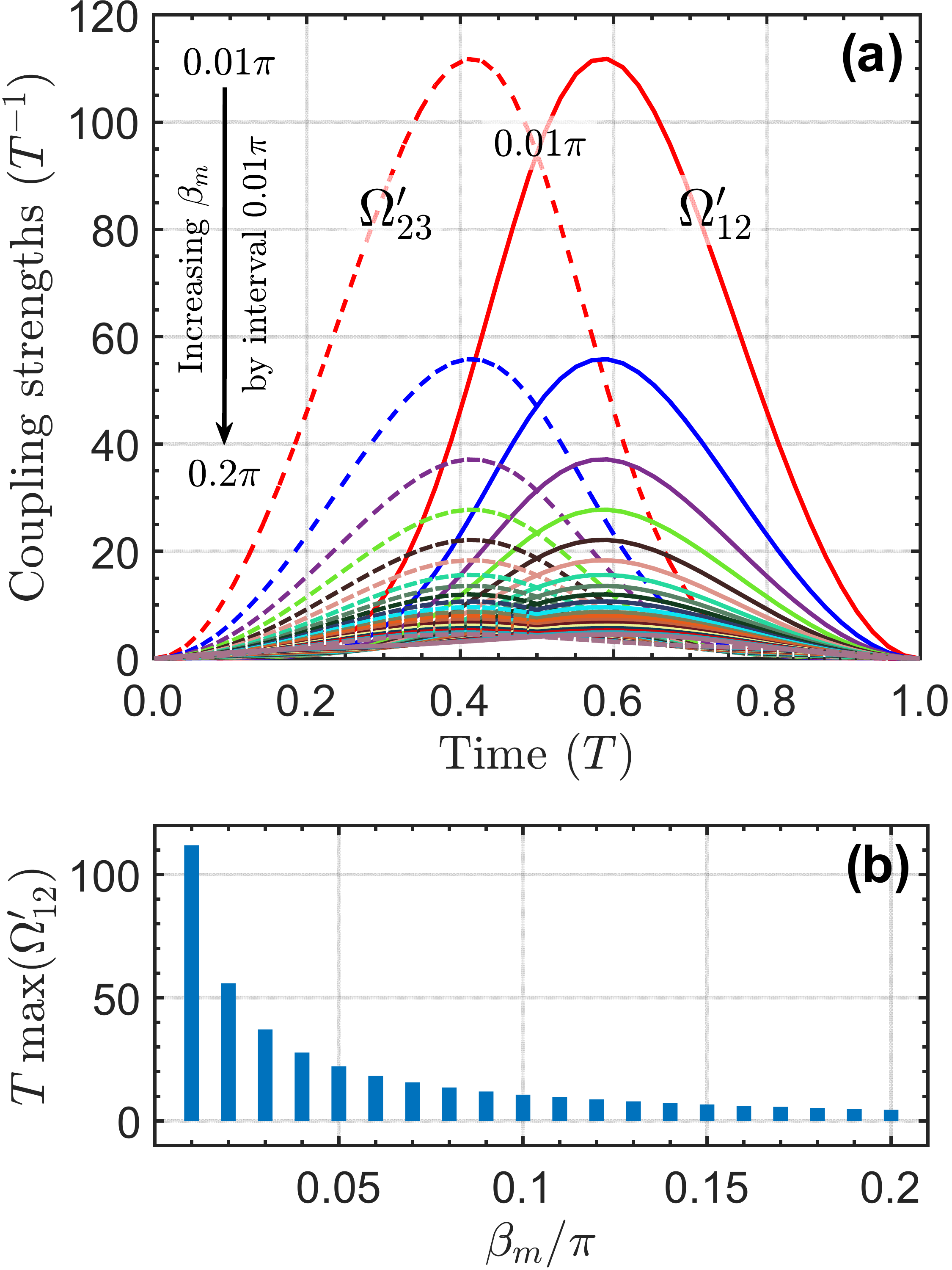}
	\caption{(a)~Time-dependent coupling strength with different values of $\beta_m$. The solid lines and dotted lines represent the driving pulses $\Omega_{12}^{\prime}(t)$ and $\Omega_{23}^{\prime}(t)$, respectively. (b)~The relation between the values of $\beta_m$ and $T$max[$\Omega_{12}^{\prime}(t)$].}\label{f9}
\end{figure}
\begin{figure*}
	\includegraphics[width=0.88\linewidth]{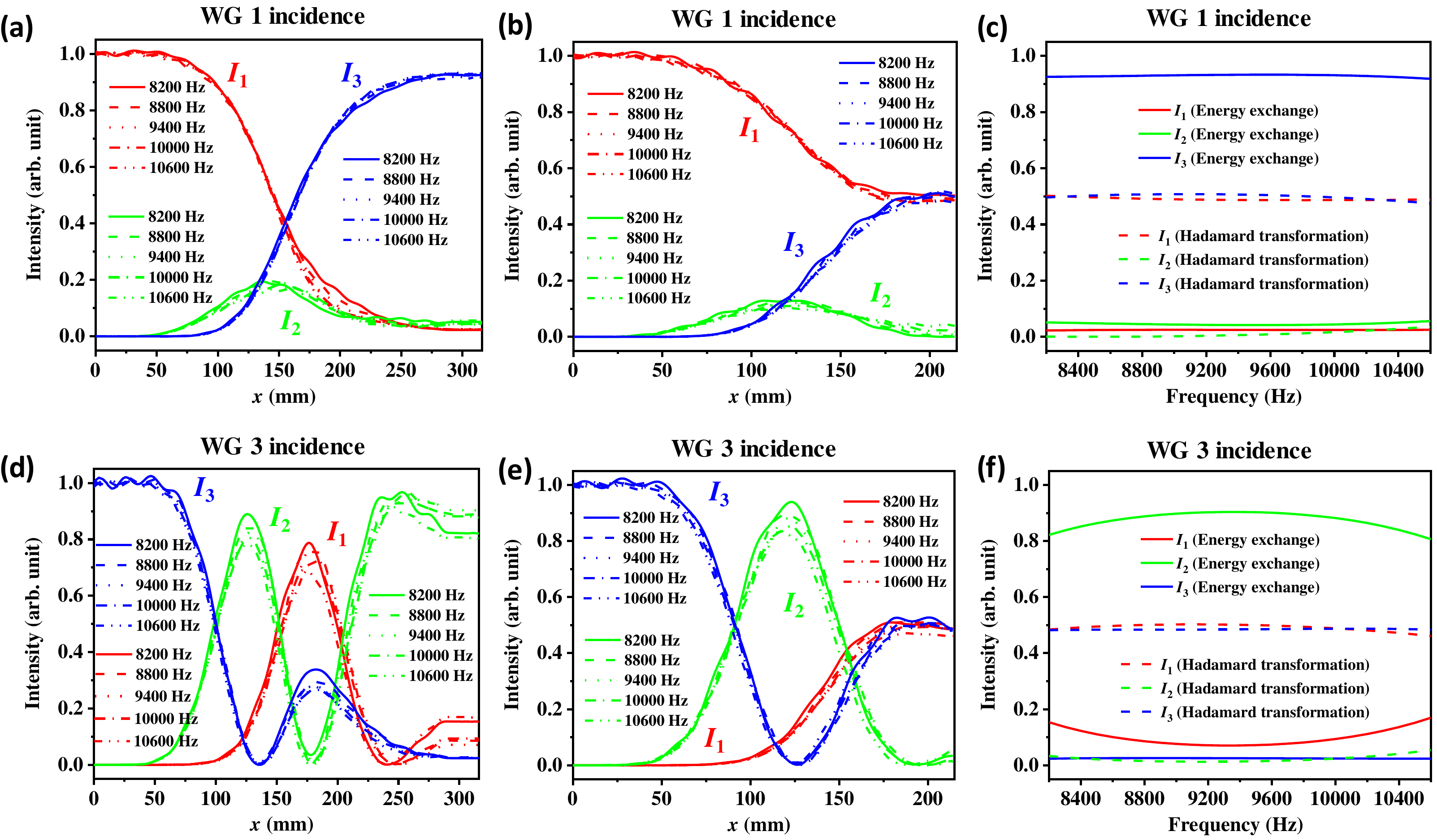}
	\caption{Acoustic intensity distributions along three WGs at 8200-10600 Hz for (a), (b) WG 1 incidence and (d), (e) WG 3 incidence. Energy exchange is realized in (a), (d), and Hadamard transformation is realized in (b), (e). The transmission spectra of three WGs within 8200-10600 Hz for (c) WG 1 incidence and (f) WG 3 incidence.}\label{f10}
\end{figure*}
\textcolor{black}{To offset nonadiabatic errors when the adiabatic criterion is not satisfied well, a correction Hamiltonian $\hat{H}_{\mathrm{c}}(t)$ is needed to modify Hamiltonian (\ref{e1}) as $\hat{H}_{\mathrm{mod}}(t)=\hat{H}(t)+\hat{H}_{\mathrm{c}}(t)$, and a unitary $\hat{V}(t)$ selecting a basis of dressed states is required to guarantee the agreement between dressed medium states and original medium states~[i.e., eigenstates of Hamiltonian (\ref{e1})] at instants $t_{i}$ and $t_{f}$. Here, the construction of dressed states is parametrized as a rotation of the spin with Euler angles $\xi(t)$, $\mu(t)$, and $\eta(t)$~\cite{Baksic2016}
\begin{equation}
\hat{V}_{g}=\exp \left[i \eta(t) \hat{M}_{z}\right] \exp \left[i \mu(t) \hat{M}_{x}\right] \exp \left[i \xi(t) \hat{M}_{z}\right].
\end{equation}
The correction Hamiltonian $\hat{H}_{\mathrm{c}}(t)$ is given by a general form as
\begin{equation}
\hat{H}_{c}(t)=\hat{U}_{\mathrm{ad}}^{\dagger}(t)\left[g_{x}(t) \hat{M}_{x}+g_{z}(t) \hat{M}_{z}\right] \hat{U}_{\mathrm{ad}}(t).
\end{equation}
which leads to a modified Hamiltonian with the same form as Hamiltonian (\ref{e1})
\begin{equation}\label{eA3}
\hat{H}_{\mathrm{mod}}=\left(\begin{array}{ccc}
0 & \Omega^m_{12}(t) & 0 \\
\Omega^m_{12}(t) & 0 & \Omega^m_{23}(t) \\
0 & \Omega^m_{23}(t) & 0
\end{array}\right),
\end{equation}
with $\Omega^m_{12}(t)=\Omega(t) \sin \theta(t)-g_{x}(t) \cos \theta(t)+g_{z}(t) \sin \theta(t)$, and $\Omega^m_{23}(t)=-\Omega(t) \cos \theta(t)-g_{z}(t) \cos \theta(t)-g_{x}(t) \sin \theta(t)$.
Meanwhile, the Hamiltonian in the new frame defined by $\hat{V}(t)$ takes the form of
\begin{eqnarray}\label{eA4}
\hat{H}_{\text {new }}(t)&=&\hat{V}(t) \hat{H}_{\mathrm{ad}}(t) \hat{V}(t)^{\dagger}+\hat{V}(t) \hat{U}(t) \hat{H}_{c}(t) \hat{U}(t)^{\dagger} \hat{V}(t)^{\dagger}\nonumber\\
&&+i \frac{d \hat{V}(t)}{d t} \hat{V}(t)^{\dagger},
\end{eqnarray}
in which the unwanted off-diagonal elements should be canceled to offset nonadiabatic errors. Thus the parameters of $g_{x}(t)$ and $g_{z}(t)$ can be determined by:
\begin{subequations}
	\begin{equation}
	g_{x}(t)=\frac{\dot{\mu}(t)}{\cos \xi(t)}-\dot{\theta}(t) \tan \xi(t),
	\end{equation}
	\begin{equation}
	g_{z}(t)=-\Omega(t)+\dot{\xi}(t)+\frac{\dot{\mu}(t) \sin \xi(t)-\dot{\theta}(t)}{\tan \mu(t) \cos \xi(t)}.
	\end{equation}
\end{subequations}
For simplicity, $\xi(t)=0$ is adopted in this work. In this situation, Hamiltonian Eq.~(\ref{eA4}) is diagonal in the representation of the following three dressed states
\begin{eqnarray}
|\Psi_{0}(t)\rangle&=&\cos \mu(t)[\cos \theta(t)|1\rangle+\sin \theta(t)|3\rangle]+i \sin \mu(t)|2\rangle,\nonumber\\
|\Psi_{\pm}(t)\rangle&=&\left\{[\sin \theta(t) \mp i \sin \mu(t) \cos \theta(t)]|1\rangle\mp \cos \mu(t)|2\rangle\right.\nonumber\\
&&\left.-[\cos \theta(t) \pm i \sin \mu(t) \sin \theta(t)]|3\rangle\right\}/{\sqrt{2}}.
\end{eqnarray}
}

\textcolor{black}{The state function of the system can be mapped into the dressed-state space by 
\begin{equation}\label{e13}
\Psi(t) =\sum_{n=\pm,0} \tilde{a}_{n}(t) \Psi_{n}(t) ,
\end{equation}
where $\tilde{a}_{n}(t)=a_{n}(t_{i}) \exp \left[-i \int_{t_{i}}^{t} \varepsilon_{n}\left(t^{\prime}\right) d t^{\prime}\right]$ with $a_{n}(t_{i})=\langle\Psi_{n}(t_{i})|\Psi(t_{i})\rangle$ and $\varepsilon_{n}(t)=\langle\Psi_{n}(t)|\hat{H}_{\mathrm{mod}}(t)|\Psi_{n}(t)\rangle$. For a given condition of $|a_{0}\left(t_{i}\right)|=1$ and $a_{+}\left(t_{i}\right)=a_{-}\left(t_{i}\right)=0$, the state function follows the evolution of $|\Psi_{0}(t)\rangle$. Perfect state transfer from $|1\rangle$ to $|3\rangle$ can be achieved when 
$\mu(t)$ and $\theta(t)$ meet the condition of 
\begin{equation}\label{eA8}
\mu\left(t_{i}\right)=\mu\left(t_{f}\right)=0,\quad
\theta\left(t_{i}\right)=0,\quad\theta\left(t_{f}\right)=\pi / 2.
\end{equation}
In this case,  the adiabatic criterion can be avoided and the same result of the final state as $\left|\psi_{0}\left(t_{f}\right)\right\rangle$ can be achieved by designing $\mu(t)$ and $\theta(t)$ meticulously, resulting in a shorter time of perfect evolution.}

\textcolor{black}{According to the analogy between two Hamiltonians (\ref{e1}) and (\ref{eA3}), it can be seen that the driving pulses in three-level system for STAP are changed to
\begin{eqnarray}
&&\Omega_{12}^{\prime}(t)=-\dot{\theta}(t) \sin \theta(t) \cot \mu(t)-\dot{\mu}(t) \cos \theta(t),\nonumber\\
&&\Omega_{23}^{\prime}(t)=\dot{\theta}(t) \cos \theta(t) \cot \mu(t)-\dot{\mu}(t) \sin \theta(t).
\end{eqnarray}
To meet the conditions in Eq.~(\ref{eA8}), we set the start and end instants $t_i=0$, $t_f=T$ and pulse parameters~\cite{kangyihaopra2016,JLWu2017}:
\begin{equation}\label{eA10}
\begin{aligned}
&\theta(t)=\frac{\pi t}{2 T}-\frac{1}{3} \sin \left(\frac{2 \pi t}{T}\right)+\frac{1}{24} \sin \left(\frac{4 \pi t}{T}\right),\\
&\mu(t)=\frac{\beta_m}{2}\left[1-\cos \left(\frac{2 \pi t}{T}\right)\right],
\end{aligned}
\end{equation}
where $\beta_m$ is a parameter related to evolutive populations of $|1\rangle$. Two driving pulses $\Omega_{12}^{\prime}(t)$ and $\Omega_{23}^{\prime}(t)$ for STAP are related to the value of $\beta_m$. Figure~\ref{f9}(a) shows the time-dependent coupling strength with different values of $\beta_m$, from which we can observe that the demand for maximum coupling strength decreases when the value of $\beta_m$ increases, offering a variety of options for the design of driving pulses. Meanwhile, from Fig.~\ref{f9}(b) we learn that a higher value of $\beta_m$ results in a shorter evolution time with a fixed maximum coupling strength, making it necessary to control the value of $\beta_m$ as high as possible for designing compact acoustic device. Nevertheless, as another constraint to determine the value of $\beta_m$, as described in $|\Psi_0(0)\rangle$, a high value of $\beta_m$ inevitably leads to a large population along mediate state $|2\rangle$. Although the lossy caused by population along intermediate state can be neglected in WG system compared with that in atomic system designed by quantum adiabatic technology, it would have influence on the one-way feature of the acoustic WG system (see Section~\ref{Sec3_3}). Therefore, a compromise consideration is needed to determine the value of $\beta_m$. For the initial time $t_{i}=0$, Eq.~(\ref{eA10}) could agree well with the condition (\ref{eA8}), providing a practical approach to STAP, so as to design compact acoustic device.}

\section{Broadband performance of acoustic three-WG couplers}\label{AppendixB}
{In our work, we construct two types of three-WG coupler based on STAP, through which energy exchange and Hadamard transformation can be realized respectively. For energy exchange and Hadamard transformation, the acoustic intensity distributions along three WGs at 8200-10600 Hz with WG 1 incidence are shown in Figs.~\ref{f10}(a) and (b), respectively, while those with WG 3 incidence are shown in Figs.~\ref{f10}(d) and (e), respectively. The propagation process within the operating band from 8200 Hz to 10600 Hz can almost remain unchanged, verifying that the proposed couplers are robust to the incident frequency. Meanwhile, the transmission spectra of three WGs within 8200 Hz-10600 Hz for WG 1 incidence and WG 3 incidence are shown in Figs.~\ref{f10}(c) and (f), respectively. The normalized output intensities for three WGs (1, 2, 3) in the cases of energy exchange and Hadamard transformation are about (0.03, 0.05, 0.92) and (0.49, 0.01, 0.50) in average for WG 1 incidence [Fig.~\ref{f10}(c)], while those are about (0.11, 0.86, 0.03) and (0.49, 0.02,0.49) in average for WG 3 incidence [Fig.~\ref{f10}(f)]. The output intensities of three WGs meet the results of energy exchange and Hadamard transformation within 8200-10600 Hz, which further confirm the broadband feature of the coupler. Therefore, as illustrated in Fig.~\ref{f8}(d)-(f), the unidirectional acoustic beam splitter composed of several couplers can work well in a continuous band within 8200-10600 Hz.} 

\bibliography{apssamp}
\end{document}